\documentclass[a4paper,english,aps,preprint]{revtex4}
\usepackage{pslatex}
\usepackage[T1]{fontenc}
\usepackage[latin1]{inputenc}
\usepackage{amsmath}
\usepackage{graphicx}
\usepackage{amssymb}

\makeatletter

\newcommand{\noun}[1]{\textsc{#1}}

\usepackage{babel}
\makeatother
\begin{document}

\preprint{Accepted for Journal of Synchrotron Radiation}

\title{Linear and Circularly Polarized Light to Study Anisotropy and Resonant
Scattering in Magnetic Thin Films}

\author{Hélio C. N. Tolentino}

\affiliation{LNLS, Laboratório Nacional de Luz Síncrotron, CP 6192, 13084-971,
Campinas, SP, Brazil}

\author{Júlio C. Cezar}

\affiliation{LNLS, Laboratório Nacional de Luz Síncrotron, CP 6192, 13084-971,
Campinas, SP, Brazil}

\affiliation{IFGW, Universidade Estadual de Campinas, CP 6165, 13083-970, Campinas,
SP, Brazil}

\author{Narcizo M. Souza-Neto}

\affiliation{LNLS, Laboratório Nacional de Luz Síncrotron, CP 6192, 13084-971,
Campinas, SP, Brazil}

\affiliation{Departamento de Física dos Materiais e Mecânica, IF-USP, São Paulo,
SP, Brazil}

\author{Aline Y. Ramos}

\affiliation{LNLS, Laboratório Nacional de Luz Síncrotron, CP 6192, 13084-971,
Campinas, SP, Brazil}

\affiliation{LMCP, Laboratoire de Minéralogie-Cristallographie, UMR 7590 , CNRS,
Paris, France}

\begin{abstract}
The remarkable polarization properties of the synchrotron light have
lead to the advent of modern synchrotron-related spectroscopic studies
with angular and/or magnetic selectivity. We give here an overview
of the prominent aspect of the polarization of the light delivered
by a bending magnet, and some dichroic properties in X-ray Absorption
Spectroscopy (XAS). We report then two studies developed at the Brazilian
Synchrotron Light Laboratory (LNLS), exemplifying the profit gained
using linear and circular polarization of the X-ray for the study
of magnetic thin films and multilayers. Angle-resolved XAS was used
in strained manganite thin films to certify a model of local distortion
limited within the $MnO_{6}$ polyhedron. A pioneer experience of
X-ray magnetic scattering at grazing incidence associated with dispersive
XAS in a $Co/Gd$ multilayer draws new perspectives for magnetic studies
in thin films and multilayers in atmospheric conditions in the hard
X-ray range. 
\end{abstract}


\maketitle

\section{Introduction}

The importance of the synchrotron light in the investigation of new
materials is well recognized nowadays. Synchrotron light sources deliver
much more photons and with a larger brilliance compared to conventional
tube sources. Furthermore, these sources owe their status to other
intrinsic properties: wide and continuous spectrum, which allows tuning
the photon energy close to resonances, and well defined polarization
states. Both aspects are of fundamental interest for the investigation
of local atomic and electronic anisotropy, and local magnetic moment
and magnetic order in magnetic systems. 

In this paper, we discuss the prominent aspects related to the use
of linear and circular polarized light emited in a bending magnet
and present, as examples of this use in thin films and multilayers,
two studies performed at the Brazilian Synchrotron Light Laboratory
(LNLS). We begin (section 2) with an introduction to synchrotron light
emission and polarization properties, followed by the description
of the optical set-up's and how they affect the polarization states.
We limit this approach to the light produced by bending magnets, once
the studies presented were performed on such devices. We introduce
(section 3) X-ray Absorption Spectroscopy (XAS) and some dichroic
properties of the spectra related to the polarization of the photons.
Examples are presented in the two next sections. The linear polarization
of the photons (section 4) has been exploited at the double-crystal
monochromator XAS beam line to study the local distortion in strained
manganite thin films. The dispersive XAS beam line, well suited for
X-ray Magnetic Circular Dichroism (XMCD) technique due to its intrinsic
stability, has been associated to grazing incidence technique to measure
(section 5) resonant scattering in a $Co/Gd$ magnetic multilayer.
These two examples highlight the potentiality of use of the polarization
properties of the synchrotron light in the XAS studies of atomic,
electronic and magnetic properties of materials.

\section{Synchrotron light sources and polarization properties}

The aim of this section is to recall some essential features of the
light emitted by relativistic electrons within the poles of a bending
magnet in a storage ring. This section is based on the development
made by J. D. Jackson \cite{Jackson-75}. The properties of the electromagnetic
radiation emitted by these electrons are related to their trajectory
and motion. The electrons revolve along a closed trajectory with quasi-circularly
bent parts, which makes the acceleration well-defined and radial.
Let us consider the special situation of a point charge in instantaneously
circular motion, so that its acceleration is perpendicular to its
velocity (Fig. 1). In a synchrotron one deals with the ultra relativistic
case where $\gamma\gg1$, with $\gamma$ defined as the energy of
the electrons divided by its rest energy $mc^{2}$. At LNLS, operating
with an electron energy of 1.37 GeV, $\gamma=2680$; for a high energy
machine, like APS in Chicago operating at 7 GeV, $\gamma=13700$. 

The propagating radiation field generated by a point accelerated charge
can be written as function of its normalized velocity $(\overrightarrow{\beta}=\frac{\overrightarrow{v}}{c})$
and acceleration $(\overrightarrow{\beta}'=\frac{d\overrightarrow{\beta}}{dt})$:

\begin{equation}
\overrightarrow{B}(\overrightarrow{r},t)=[\hat{n}\times\overrightarrow{E}]_{ret}\end{equation}

\begin{equation}
\overrightarrow{E}(\overrightarrow{r},t)=\frac{e}{c}\left[\frac{\hat{n}\times\{(\hat{n}-\overrightarrow{\beta})\times\overrightarrow{\beta'}\}}{(1-\overrightarrow{\beta}\cdot\hat{n})^{3}R}\right]_{ret}\end{equation}

where both $\overrightarrow{E}$ and $\overrightarrow{B}$, evaluated
at the retarded time, are transverse to the radius vector $\hat{n}$
and decay as $1/R$, the distance from the source to the observer.
The Poynting`s vector $[\overrightarrow{S}.\hat{n}]_{ret}=\frac{c}{4\pi}\left|\overrightarrow{E}\right|^{2}$
gives the density of propagating energy.

\begin{figure}
\begin{center}\includegraphics{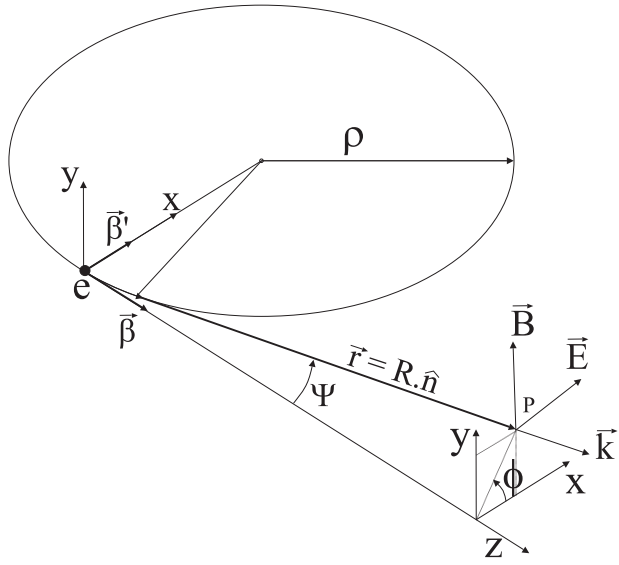}\end{center}

\caption{Geometry and notation used to calculate the synchrotron radiation
emission.}
\end{figure}

The specific spatial relationship between $\overrightarrow{\beta}$
and $\overrightarrow{\beta'}$ determines the detailed angular distribution.
However, the relativistic effect, represented by the presence of the
factors $(1-\overrightarrow{\beta}.\hat{n})$ in the denominator,
dominates the whole angular distribution and gives rise to the collimation
of the emission. By integrating the energy per unit area per unit
time during a finite period of acceleration, one can deduce the power
radiated per unit solid angle. 

\begin{equation}
\frac{dP(t')}{d\Omega}\simeq\frac{2e^{2}}{\pi c^{2}}\frac{\gamma^{6}\beta^{2}}{(1+\gamma^{2}\Psi^{2})^{3}}\left[1-\frac{4\gamma^{2}\Psi^{2}cos^{2}\phi}{(1+\gamma^{2}\Psi^{2})^{2}}\right]\end{equation}
In the ultra relativistic approximation $(\gamma\gg1)$, the root
mean square angle of emission $\left\langle \Psi^{2}\right\rangle ^{1/2}$
is given by $\gamma^{-1}$. The emission is essentially within a cone
with a very small opening about the orbit plane. On the opposite situation,
for an accelerated charge in non-relativistic motion, the angular
distribution of the emitted radiation shows a simple $sin^{2}\varphi$
behavior, where $\varphi$ is measured with respect to the direction
of acceleration. 

\begin{figure}
\begin{center}\includegraphics{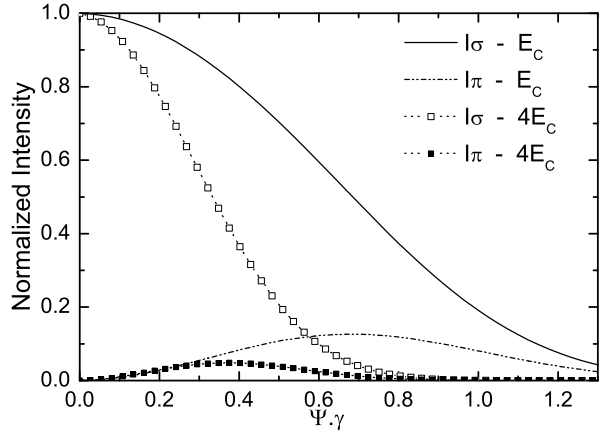}\end{center}

\caption{Normalized intensity calculated for the emitted radiation with the
polarization vector parallel ($\sigma$) and perpendicular ($\pi$)
to the electron orbit plane, as function of the displacement angle
from this plane ($\Psi\gamma$), at the energies $E_{C}=2080eV$ (lines)
and $4E_{C}=8320eV$ (symbols). For the LNLS storage ring, $\gamma^{-1}=0.373mrad$.}
\end{figure}

The total power radiated can be found by integrating the above equation
over all angles

\begin{equation}
P(t')=\frac{2}{3}\frac{e^{2}\beta'^{2}}{c^{2}}\gamma^{4}\end{equation}

This power has a broad band in the frequency space that comes as a
counterpart of the collimation. An observer, who sits in the laboratory,
far from the moving point charge, sees the emission only when the
charge travels along a very specific portion of the circular trajectory,
limited to about $1/\gamma$, and receives light as short pulses.
The Fourier transform of the delta-like function in time space is
spread over a wide range in the frequency space, so that, the synchrotron
emission has a continuous \char`\"{}white light\char`\"{} distribution
\cite{Raoux-Hercules93}.

The vectorial double cross product in the transverse electric field
generated by the relativistic moving charge gives a vector with two
components, one along $\hat{n}-\overrightarrow{\beta}$ and the other
along the acceleration $\overrightarrow{\beta}'$:\begin{equation}
\hat{n}\times\{(\hat{n}-\overrightarrow{\beta})\times\overrightarrow{\beta}'\}=(\hat{n}-\overrightarrow{\beta})(\hat{n}\cdot\overrightarrow{\beta}')-\overrightarrow{\beta}'(1-\hat{n}\cdot\overrightarrow{\beta})\end{equation}

$\hat{n}-\overrightarrow{\beta}$ is roughly a difference between
two almost collinear unit vectors, therefore it is orthogonal to $\hat{n}$.
If one considers the emission right above the plane of the circular
orbit $(\phi=\pi/2)$, $\hat{n}-\overrightarrow{\beta}$ will be a
vector perpendicular to the orbit plane and $\overrightarrow{\beta}'$
a vector in this plane. Since the acceleration is the derivative of
the velocity, one of the components of the electric field is phase
shifted by $\pi/2$ with respect to the other. This is at the origin
of the circular polarized light. 

The energy radiated per unit frequency interval per unit solid angle
is given by

\begin{equation}
\frac{d^{2}I}{d\omega d\Omega}=\frac{3e^{2}}{4\pi^{2}c}\left(\frac{\omega}{\omega_{c}}\right)^{2}\gamma^{6}\left(\frac{1}{\gamma^{2}}+\Psi^{2}\right)^{2}\left[K_{2/3}^{2}(\xi)+\frac{\Psi^{2}}{(1/\gamma^{2})+\Psi^{2}}K_{1/3}^{2}(\xi)\right]\end{equation}
where $K_{1/3}(\xi)$ and $K_{2/3}(\xi)$ are second rank Bessel functions,
with $\xi=\frac{\omega\rho}{3c}\left(\frac{1}{\gamma^{2}}+\Psi^{2}\right)^{3/2}$.
The parameter $\omega_{c}=\frac{3c}{2\rho}\gamma^{3}$ defines the
critical energy, $E_{c}=\hbar\omega_{c}$, that separates the total
emission in two equal parts. At LNLS, $E_{c}$ is 2.08 keV, characterizing
a low energy storage ring \cite{Rodrigues-JSR98}. The first term
in square bracket corresponds to radiation polarized parallel to the
orbit and the second to radiation polarized perpendicular to that
plane. For $\Psi=0$ the photons are strictly linearly polarized in
the plane. Figure 2 shows the calculated intensity of the components
parallel $(\sigma)$ and perpendicular $(\pi)$ as function of the
displacement angle $(\Psi)$, at the critical energy $E_{c}$ and
at $4E_{c}$, energy close to the working range of the XAS experiments.
One can remark that in the high energy region the cone of emission
gets narrower.

The degree of linear polarization in the horizontal plane $P_{L}$
and of circular polarization $P_{C}$ are given by\[
\begin{array}{ccccc}
P_{L}=\frac{A^{2}-B^{2}}{A^{2}+B^{2}} &  & ; &  & P_{C}=\frac{2AB}{A^{2}+B^{2}}\end{array}\]

With \begin{equation}
\begin{array}{ccccc}
A=\left(\frac{1}{\gamma^{2}}+\Psi^{2}\right)K_{2/3}(\xi) &  & ; &  & B=\left(\frac{1}{\gamma^{2}}+\Psi^{2}\right)^{1/2}\Psi K_{1/3}(\xi)\end{array}\end{equation}
 Circularly polarized light can then be obtained by selecting the
incident beam slightly above or below the orbital plane. 

We limited this section to the polarization properties of the light
produced in a bending magnet, once our purpose is to give the basic
principles of the emission and also because this was the source used
in the studies developed at the LNLS. Circularly polarized light are
actually more commonly obtained by using special insertion devices
like wiggler or undulators \cite{Raoux-Hercules93,Nielsen-01}. To
a lesser extent, optical devices like quarter-wave plates are used
to transform the linearly polarized light in the orbital plane into
circularly polarized light \cite{Giles-JAC94}. 

%\onecolumn%
\begin{figure}
\noindent \begin{center}\includegraphics[%
  scale=0.6]{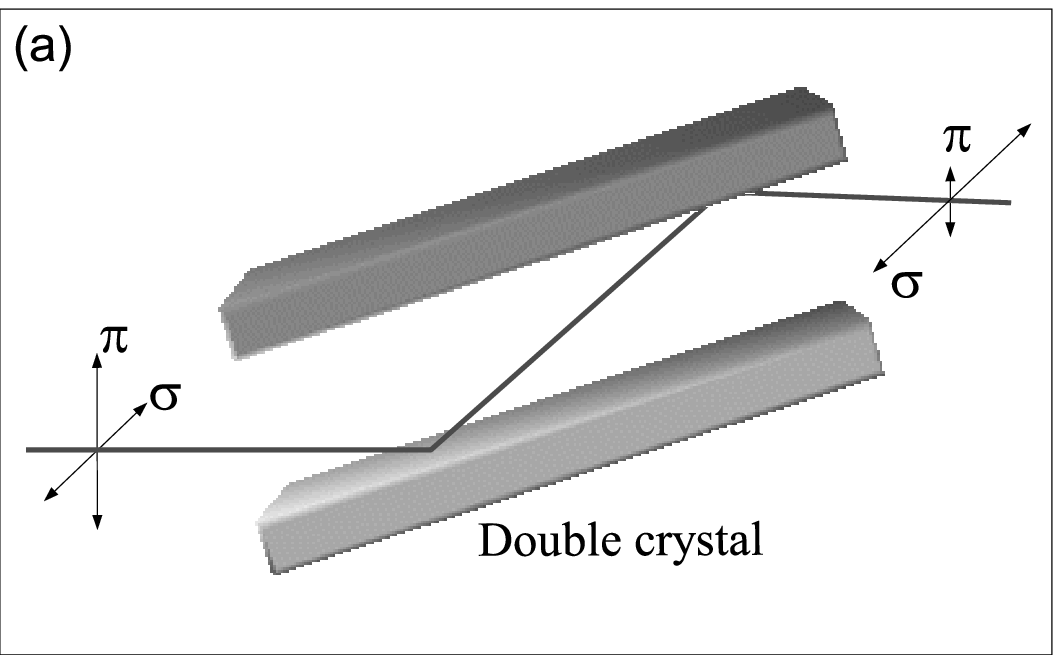}\includegraphics[%
  scale=0.6]{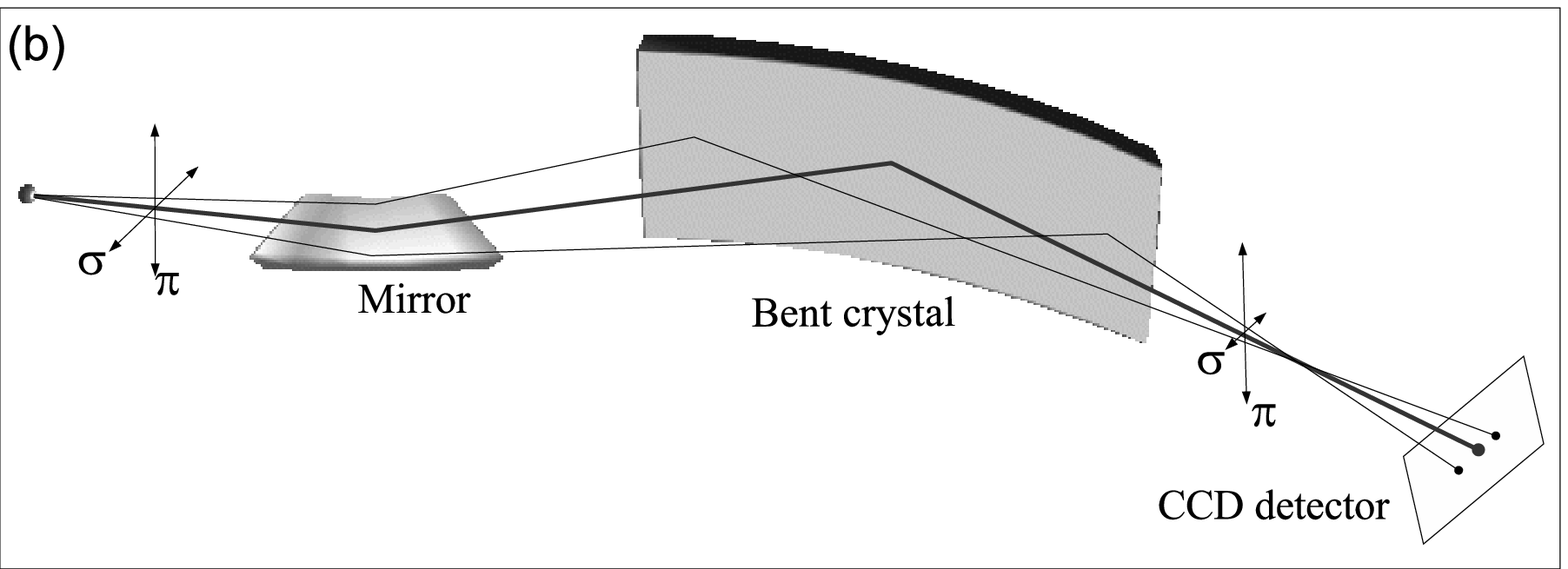}\end{center}

\caption{Schematic view of the optics for (a) the conventional double reflection
monochromator, and (b) the dispersive monochromator, where an energy
band-pass is selected by a bent crystal. }
\end{figure}
%\twocolumn

In a real synchrotron, the average over the finite slit aperture and
the electron emittance, which is roughly the product of the electron
beam size by its divergence, attenuate the degree of polarization
by a few percents. The effects of the optical elements are much larger
and only these effects are considered in the scope of this paper.
Mirrors at grazing incidence reflect both components of the beam in
the same way and the effect on polarization ratio can always be neglected.
Conventional XAS monochromator involves two reflections from silicon
or germanium crystals with the horizontal $(\sigma)$ component of
the beam lying perpendicular to the diffraction plane (Fig.3a). For
this component, the polarization factor in the reflected beam is equal
to one and, neglecting absorption, the beam is totally reflected within
the Darwin width \cite{James-65}. The vertical $(\pi)$ component
of the beam lies in the diffraction plane and is reduced according
to the factor $\left|cos2\theta\right|^{2}$, with $\theta$ the Bragg
angle of diffraction. The total emitted power is almost unchanged
(Fig. 4a) and the linear polarization is enhanced (Fig. 4b). In the
case of the dispersive XAS monochromator \cite{Dartyge-NIM86}(Fig.
3b), there is only one Bragg diffraction. The horizontal $(\sigma)$
component lying in the diffraction plane is affected by the factor
$\left|cos2\theta\right|$. The vertical $(\pi)$ component is not
affected. The total emitted power is reduced (Fig. 4a), but circular
polarization is significantly enhanced (Fig. 4c). 

\begin{figure}
\begin{center}\includegraphics{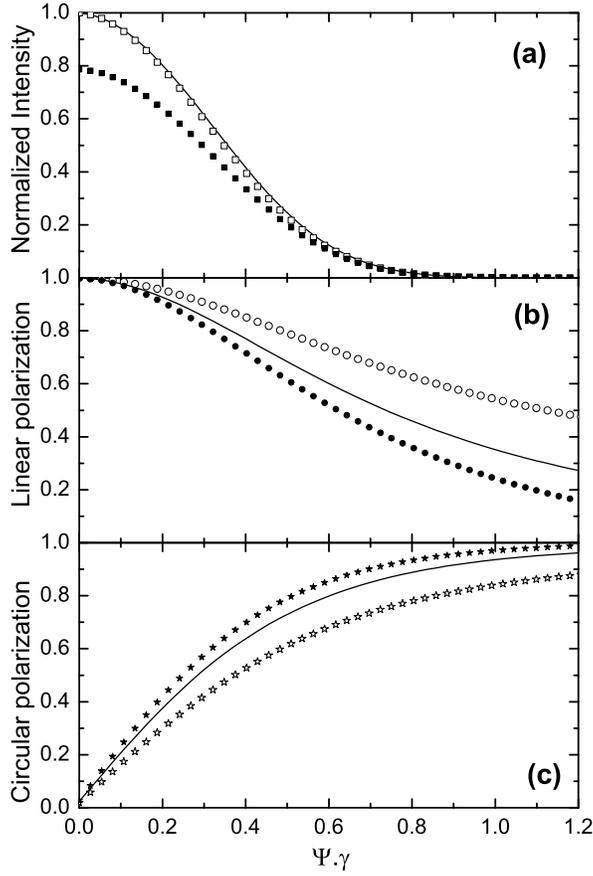}\end{center}

\caption{Normalized intensity (a) linear polarization (b) and circular polarization
(c) ratios at $Gd$ $L_{3}$ edge$(E=7243eV)$, calculated for the
photon beam produced by the source (solid line) and after diffraction
by the dispersive (closed symbols) and the double-crystal monochromator
(open symbols), as function of the displacement angle ($\Psi$) from
the electron orbit plane. }
\end{figure}

\section{Polarized Light and X-ray Absorption Spectroscopy}

\subsection{\noun{X-ray absorption spectroscopy }}

An X-ray absorption spectrum corresponds to the variation of the absorbing
coefficient with the energy of the photons\cite{Koningsberger-88}.
The absorption edge is the drastic variation of this absorption coefficient
for a photon energy close to the energy of a core level electron.
It corresponds to the transition of deep core electron level of a
selected atom to empty levels above the Fermi energy. The ratio between
the absorbed energy and the incident photon flux, called absorption
cross section, $\sigma(\omega)$ is given by the summation over all
possible final states of the transition probability from the initial
state $i$ to a final state $f$. Each probability is expressed by
the Fermi's golden rule as the square matrix related to the Hamiltonian
describing the interaction of X-rays with the electrons of the matter.
This interaction Hamiltonian can be approximated in the perturbation
theory and expressed as a multipolar expansion. The first term of
the expansion is the dipole electric term, followed by the dipole
magnetic and the quadrupole electric terms, several orders of magnitude
smaller \cite{Fontaine-Hercules93}. In the so called electric dipole
approximation, these last terms are neglected and the electromagnetic
field is taken as constant in the initial state site. The absorption
cross section then reads\begin{equation}
\sigma(\omega)=4\pi²\alpha\hslash\omega\sum_{if}\left|\left\langle f\left|\hat{\epsilon}\cdot\vec{r}\right|i\right\rangle \right|^{2}\delta(E_{f}-E_{i}-\hbar\omega)\rho(E_{f}),\end{equation}

where $\alpha\equiv e²/(\hslash c)$ is the fine structure constant.
The delta Dirac distribution ensures the energy conservation and $\rho(E_{f})$
is the empty density of states (DOS). 

The chemical selectivity is the main appeal of the X-ray absorption
spectroscopy. Each element of the material can be selected by tuning
the X-ray photon energy in the region of one of its core levels and
experimental data can be collected element by element. 

The final and initial states in the matrix cross section can be expressed
as a combination of spherical harmonic and their composition obey
the Wigner-Eckard theorem. The resulting selection rules are the second
major characteristic of X-ray absorption spectroscopy. In the dipole
approximation these selection rules state $\Delta s=0$, $\Delta l=\pm1$,
$\Delta j=\pm1,0$. At the $K$ and $L{}_{\textrm{1}}$ ($l$=0) edges,
$1s$ or $2s$ electrons are excited to final empty levels with $p$
symmetry ($l$=1) and the transition to $d$ levels are forbidden
($\Delta l=2$). At the $L{}_{\textrm{2,3}}$ edges a $2p$ electron
(l=1, j=3/2, 1/2) is excited to vacant $d$ levels ($\Delta l=1$)
and, with a much weaker transition probability to empty $s$ levels
($\Delta l=-1$). X-ray absorption is then a selective probe for the
angular momentum of empty levels. 

The selection rules relative to the principal, orbital and total quantum
number do not depend on the polarization of the interacting photon.
However, the Wigner-Eckard theorem forces a last selection rule to
the magnetic orbital quantum number $m$, and this rule depends on
the polarization of the photons. If we consider non-magnetic systems
and linear polarization only, the dipole operator does not affect
the magnetic orbital quantum number and the selection rule is $\Delta m=0$
. If circularly polarized photons are used, this rule should be modified
to take into account the helicity of the photons. The modifications
of this selection rule are at the base of the interpretation of the
X- ray magnetic dichroism that will be discussed in section \ref{sub:X-ray-magnetic-dichroism} 

In summary, X-ray absorption spectra contain a bunch of information
about the ground state of the selected element in a material: local
symmetry, oxidation and spin states, spin-orbit coupling in the $2p$
and $3d$ orbitals, crystal field, covalence and charge transfer.
As a matter of fact, in the case of $3d$ transition metals essentially
structural information is obtained from the $K$ edges, while more
magnetic and electronic information is usually deduced from $L_{2,3}$
edge. 

Interpretation of X-ray absorption data can be done using basically
two approaches. The monoeletronic approach is used in band structure
or multiple scattering calculations. Except the electron which has
absorbed the photon, all electrons of the system are supposed to remain
passive in the absorption process, and the correlations among electrons
are neglected \cite{Natoli-PRA80} or taken as an overall reduction
term \cite{Rehr-RMP00}. These approximations oversimplify the description
of the X-ray absorption spectra in the case of localized final states.
This is the case of $L_{2,3}$ edge of transition metal elements,
that requires a multi-electronic approach \cite{DeGroot-JESRP94}.
Nevertheless, monoelectronic models give a good description of the
delocalized final states for systems where the electrons are weakly
correlated. They are adequate for many studies in the hard X-ray range
($0.5<\lambda<5\textrm{Å}$) and specially for the examples presented
in this paper.\\

\subsection{\noun{Linear dichroism in x-ray absorption spectroscopy}}

The polarization dependence of the intensity of the X-ray absorption
spectra can be derived quite generally from the angular dependence
of the transition matrix elements involved in the expression of the
X-ray cross section. A thorough account of the angular dependence
of the X-ray absorption spectra, including dipole and quadrupole effects
has been given by Brouder several years ago \cite{Brouder-JPCM90}.
Detailed formula to analyze the experimental spectra can be found
in this reference and we reduce the scope of this paragraph to the
expression of the simple concepts.

Let us assume that the X-ray are linearly polarized light, $\hat{\varepsilon}=\hat{x}$,
and remind that in the dipole electric approximation the square matrix
element can be written $\left|\left\langle f\left|\hat{\epsilon}\cdot\vec{r}\right|i\right\rangle \right|^{2}$.
The case of $K$ (or $L_{1}$) shell excitation is easy to visualize.
The initial $s$ state is isotropic. The vector matrix elements points
in the same direction as the $p-$component of the final state orbital
on the excited atom and thus the polarization dependence of the total
matrix element can be expressed as a function of the angle $\theta$
between the direction of the electric field vector and the direction
of the largest amplitude of the final state orbital, that is the direction
of the bond length. We should note that the $K$-shell excitations
exhibit the strongest polarization dependent effects because of the
directionality of the $p$ component in the final state, and it has
been, consequently, the earliest stated and the more widely studied
\cite{Stern-PRB74}. The case of $L_{2}$ and $L_{3}$ excitations
needs a little more mathematical treatment but elegant derivations
have also been documented \cite{Heald-77,Haskel-Phd98}.

In anisotropic media the polarization dependence of the absorption
cross section has the same structure as the dielectric constant. It
is properly described by a tensor of rank two, whose expression depends
on the point group of the media. For samples with cubic point symmetry
the absorption cross section is isotropic. The simplest expression
of the dichroic effect is obtained for non-cubic samples, with a rotation
axis of order greater than two, where one can find two different cross
sections. For this particular case, and expressing in terms of the
linear absorption coefficient $\mu\propto\sigma$, one defines two
parameters: $\mu_{\Vert}$ stands for the absorption coefficient when
the electric vector lies in a plane orthogonal to the rotation axis
and $\mu_{\bot}$ is the coefficient when the electric vector is along
the rotation axis. For any given orientation of the electric vector,
measured by the angle $\theta$ related to that rotation axis, $\mu(\theta)$
reads:\begin{equation}
\mu(\theta)=\mu_{\Vert}sin^{2}\theta+\mu_{\bot}cos^{2}\theta\end{equation}

We should note that this expression, largely known and used, is valid
over the whole XAS energy range. As it is only related to the full
point group symmetry of the material, it does not depend on the type
$(K,L,M)$ of X-ray excitation. One can readily realize that selective
pieces of information can be extracted using angle-resolved XAS, as
long as oriented samples are available. This tool, originating from
the dot product $(\hat{\epsilon}.\vec{r})$ in the cross section,
enhances the sensitivity of XAS to probe very tiny difference in anisotropic
systems, like surfaces \cite{Magnan-PRL91}, anisotropic single-crystals
\cite{Tolentino-PhyC92,Gaudry-PRB03}, multilayers \cite{Pizzini-PRBRC92}
or oriented films \cite{Souza-Neto-APL03}.

\subsection{\noun{\label{sub:X-ray-magnetic-dichroism}X-ray magnetic dichroism}}

X-ray magnetic circular dichroism (XMCD) in X-ray absorption has been
shown to be a unique element selective magnetic probe \cite{Ebert-96}.
The origin of XMCD is a local anisotropy of the absorbing atom resulting
from the expected value of the local magnetic moment $\left\langle \overrightarrow{m}\right\rangle $.
The magnetic field breaks the local symmetry of the absorber and lifts
the degeneracy of the Zeeman energy levels. Then the photoelectron
transition depends on the helicity of the photon polarization. The
XMCD signal is the difference between the absorption cross sections
with circularly left and right polarization. 

If the X-ray absorption experiment is performed using circularly polarized
photons, the selection rule for the magnetic orbital quantum number
takes into account the helicity of the photons : $\Delta m=\pm1$.
$\Delta m=+1$ for left-handed polarization and $\Delta m=-1$ for
right-handed polarization. In a magnetic compound the XMCD signal
originates from the difference in the population of the levels with
$m$ and $-m$ magnetic numbers. We recall that the interaction Hamiltonian
does not act on the spin and XMCD results from the interaction between
spin and orbital momentum, which couples the spin and the real space.
In a simplified view, the XMCD signal is expressed as $R_{XMCD}=P_{C}P_{e}\Delta\rho/\rho$,
where $\Delta\rho/\rho$ is the normalized spin polarized density
of states \cite{Schutz-87}. $P_{C}$ is the circular polarization
ratio and $P_{e}$ a Fano-like factor giving the probability for the
excited electron to be spin polarized. In the atomic limit, typical
values of $P_{e}$ are 0.01 for $K$ excitations, -0.5 for $L_{2}$
and 0.25 for $L_{3}$ excitations. This simple model has proved to
be useful for the interpretation of a large number of systems with
rather delocalized final states \cite{Baudelet-PRB91,Giorgetti-PRB93}.

When the Hamiltonian is reduced to the electric dipole term, reversing
the magnetic field or the circular polarization yield to the same
result and most of the experiments are performed with a constant helicity
with a magnetic field parallel or anti-parallel to the X-ray beam.

As XMCD, X-ray resonant magnetic scattering (XRMS) probes the spin-polarized
density of empty states above the Fermi level. In a quantum mechanical
picture, the photon excites to a higher unnoccupied level a core electron
that decays back to the initial state emitting a photon of the same
energy. The description of this process involves, then, emission via
an intermediate state and requires second-order perturbation theory.
Sensitivity to magnetism arises from the Pauli exclusion principle
and the exchange-induced splitting of the Fermi level and is reinforced
by the strong enhancement of magnetic effects when the photon energy
is close to an absorption edge \cite{Gibbs-PRL88,Hannon-PRL88}. The
XRMS signal is measured by the intensity asymmetry ratio $R_{XRMS}=(I^{+}-I^{-})/(I^{+}+I^{-})$,
where $I^{+}$ and $I^{-}$ are the intensities scattered for the
opposite directions of an applied magnetic field.

\section{Local anisotropy in perovskite thin films}

The remarkable properties of hole doped manganites show drastic sensitivity
to small changes in their structural parameters and the form of the
samples \cite{Fontcuberta-PRL96,Millis-N98,Salamon-RMP01}. The magnetism
and transport properties of manganites films differ significantly
from those of the bulk material and are dependent of the film thickness
\cite{Millis-JAP98,Prellier-JPCM01}. Such characteristic has been
associated to the strain induced by the lattice mismatch. The understanding
of the effects of the strains on the local structural parameters is
crucial to explain the behavior of manganite thin films and multilayers. 

\begin{figure}
\begin{center}\includegraphics{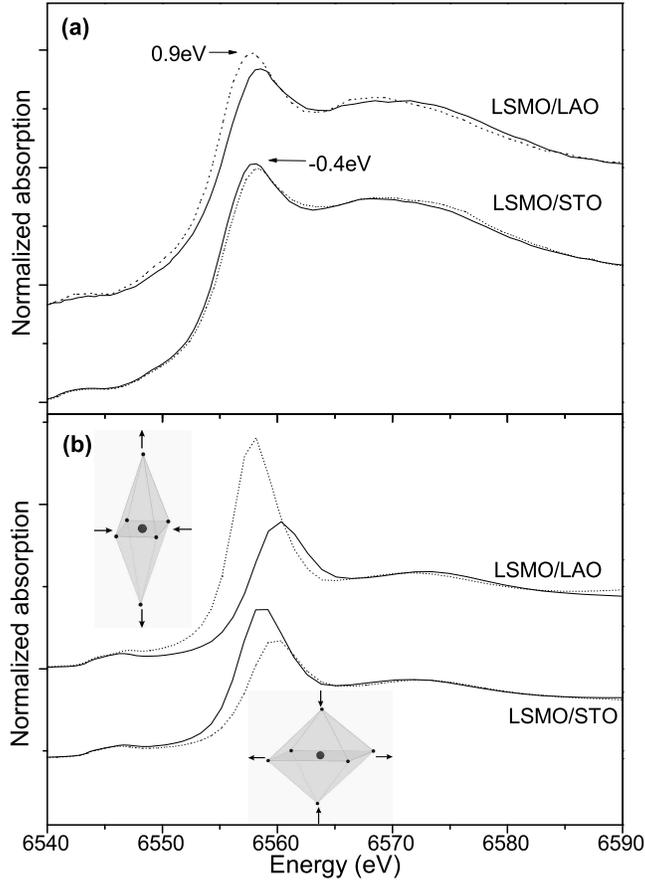}\end{center}

\caption{Experimental XAS spectra at the $Mn$ $K$ edge for the tensile ($SrTiO_{3}$)
and compressive ($LaAlO_{3}$) LSMO films in plane (solid line) and
perpendicular to film plane (dots line). The energy shifts are in
opposite directions and scaled in amplitude by a factor of two. \emph{Ab
initio} XAS calculations considering tetragonal distortion of the
$MnO_{6}$ octahedron and for the in-plane and out-of-plane contributions.}
\end{figure}

We used angle-resolved XAS to characterize the local atomic structure
around the manganese ions in $La_{0.7}Sr_{0.3}MnO_{3}$ (LSMO) thin
films epitaxially grown by pulsed laser deposition over $SrTiO_{3}$
(STO) and compressive $LaAlO_{3}$ (LAO) substrates. The small lattice
mismatch between LSMO and STO and LAO allows a pseudomorphic growth
and almost fully constrained conditions for film thickness below 100nm
and 40nm respectively \cite{Ranno-ASS02}: 1\% in-plane tensile strain
for STO and 2\% in-plane compressive strain for LAO substrate.

We were looking for a dichroic signal around manganese to characterize
a possible local anisotropy induced by the substrate. We took full
benefit of the linear polarization of the synchrotron light to perform
angle-resolved measurements and probe selectively atomic bond lengths
in the plane and out of the plane of the films. According to the Natoli's
rule \cite{Natoli-84}, the energy position of the $1s\rightarrow4p$
excitation scales as the bond length squared $(E-E_{0}\propto R^{2})$,
where $E_{0}$ is an energy close to the onset of the edge. It is
possible to follow bond length changes by carefully looking at the
energy position, that is accurately obtained thanks to the sharp rising
edge. 

Angle-resolved X-ray absorption experiments were performed at the
D04B-XAS beam line of the LNLS \cite{Tolentino-JSR01} at the Mn K-edge
(6539eV). The monochromator was a Si (111) channel-cut. An ion chamber
monitored the incident beam and the data were collected in the fluorescence
mode using a Ge 15-element solid state detector. 0.5 mm-slits selected
beam in the orbit plane $(\Psi=0)$with an acceptance of 0.03 mrad.
The light on the sample was predicted to be more than 99\% linearly
polarized. The films were set on a goniometer with the rotation axis
perpendicular to the orbit plane. The experimental spectra were taken
in two geometries: with the electric field of the incident linearly
polarized photon beam set approximately parallel $(\mu_{\Vert})$
and perpendicular $(\mu_{\bot})$ to the film surface. XAS spectra
were collected in the near-edge range of 6440 to 6700 eV with energy
steps of 0.3 eV. The energy calibration of the edge was carefully
monitored by simultaneous comparison with a Mn metal foil reference
and the spectra were normalized at about 150 eV above the edge. This
allows the edge structure in all experimental spectra to be compared
in position and intensity, and energy shifts as small as 0.1 eV are
certified.

The experimental XAS spectra of the tensile (STO substrate) and the
compressive (LAO substrate) strained films are shown in figure 5a.
Taking the out-of-plane $(\mu_{\bot})$ spectra as reference, we observe
energy shifts of the main $1s\rightarrow4p$ transition among spectra
collected in the two orientations: the shift is -0.4 eV for the tensile
film and +0.9 eV for compressive one. This result shows that in the
LSMO/STO film the average $Mn-O$ bond length in the film surface
direction is greater than the same bond distance in the perpendicular
direction to film surface. In the opposite case, in the LSMO/LAO film,
this in-plane average bond length is smaller than the out-of-plane
bond distance, with a scale factor of approximately two in the energy
shift. 

From the X-ray diffraction results on these samples \cite{Ranno-ASS02},
and following the Natoli's rule, energy shifts in opposite directions
for tensile and compressive films were expected. We should note, moreover,
that the ratio of the amplitude of the energy shifts (1:2) is approximately
the same as the ratio of the long range strain factor (\textasciitilde{}1\%
for tensile, and \textasciitilde{}2\% for compressive) among these
films. This direct proportionality between the modifications on the
cell parameters and the local octahedral modifications indicates that
the strain should be fully accommodated by changes in the coordination
polyhedra ($MnO_{6}$), without any modification of the $Mn-O-Mn$
angle \cite{SouzaNeto-PRB04}. 

\textit{Ab initio} self-consistent field calculations in the full
multiple scattering formalism \cite{Ankudinov-PRB02} were performed
to address more precisely the actual consequence of the distortion
of the octahedron on the near-edge XAS spectra and specially to investigate
how far local distortion may account for the difference observed experimentally.
The definition of the electric field vector polarization in relation
with the atomic structure orientation, and then all the photoelectron
scattering terms, are weighted using the equation 9 presented in the
section 3. This procedure enables to calculate independently the information
from the different angular contributions to the absorption. Simulations
were performed in a LSMO cluster considering isotropic $MnO_{6}$
and anisotropic octahedral distortions. As expected for the isotropic
case, the spectra do not show any modification in position and shape
of the edge main line among in-plane and out-of-plane situations.
The calculations shown in figure 5b were performed for 21-atoms cluster
with tetragonal distortion, using local order parameters scaling with
the crystallographic cell parameters of the films \cite{Souza-Neto-JAllC04}.
The calculated structures reproduce well the main features of the
experimental results. They account as well for the energy shift, in
amplitude and direction, as for the relative reduction in amplitude
of the main peak close to the edge, among the two orientations and
for each film \cite{SouzaNeto-PS04}. Based on these calculations,
we validate the model of tetragonal distortion of the $MnO_{6}$ octahedron,
suggested by experimental XANES spectra. This model is in agreement
with Extended X-ray Absorption Fine Structure results \cite{Ramos-ROMA02,SouzaNeto-PRB04}
and lead to the conclusion that the tilt angle Mn-O-Mn among adjacent
octahedra is not - or little- changed \cite{SouzaNeto-PRB04}. Actually
the model of tetragonal distortion is not the unique model that may
account for the experimental data. Orthorhombic or slighly monoclinic
distortions can not be totally ruled out only from XANES data \cite{Souza-Neto-APL03}.
However the most important result remains unchanged : the coordination
octahedron accomodated most of the distortion induced by the strain.
We conclude that this distortion, tending to localize the charge carriers,
is be the driving parameter in the modifications of the magnetic and
transport properties observed in thin films with respect to bulk systems
\cite{Souza-Neto-APL03,SouzaNeto-PRB04}.

\section{Resonant scattering from films and multilayers}

During the last decade, with the dramatic increase of the technological
interest devoted to magnetic thin films and multilayers, resonant
scattering in the soft X-ray range has progressively raised as a powerful
tool specially adapted to the study of this class of material \cite{Kao-PRL90,Tonnerre-PRL95,Sacchi-PRL98}.
However, it is worth noting that, during the same period, very few
experiments were attempted in the hard X-ray domain \cite{Dartyge-PL86}
in spite of the great advantage arising from the possibility of working
under a variety of extreme conditions. Examples of the application
of hard-X-ray XRMS can be found only recently the literature with
the works of Seve et al. and Jaouen et al. on the $5d$ electronic
states across the rare earth layers in $Ce/Fe$ and $La/Fe$ multilayers
\cite{Seve-PRB99,Jaouen-PRB02}. 

\begin{figure}
\begin{center}\includegraphics{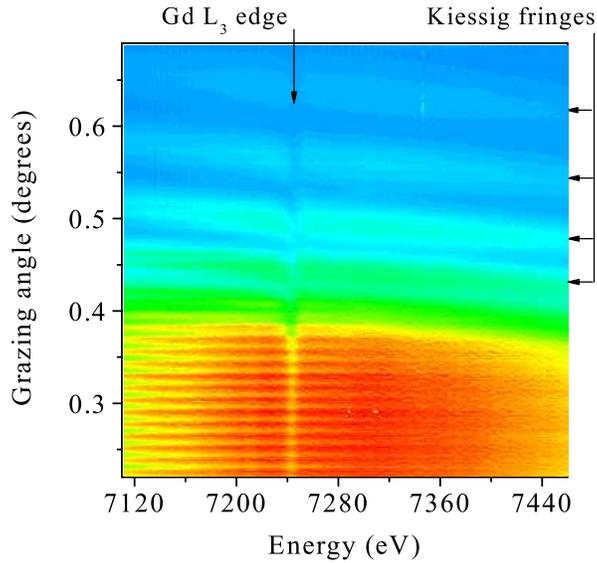}\end{center}

\caption{X-ray resonant scattering intensity around the $Gd$ $L_{3}$ edge
for the multilayer structure $(Co_{1nm}/Gd_{0.2nm})_{40}$ over $SiO_{2}$.
The edge is indicated by the vertical arrow. Kiessig fringes are also
indicated by lateral arrows. }
\end{figure}

\begin{figure}
\begin{center}\includegraphics{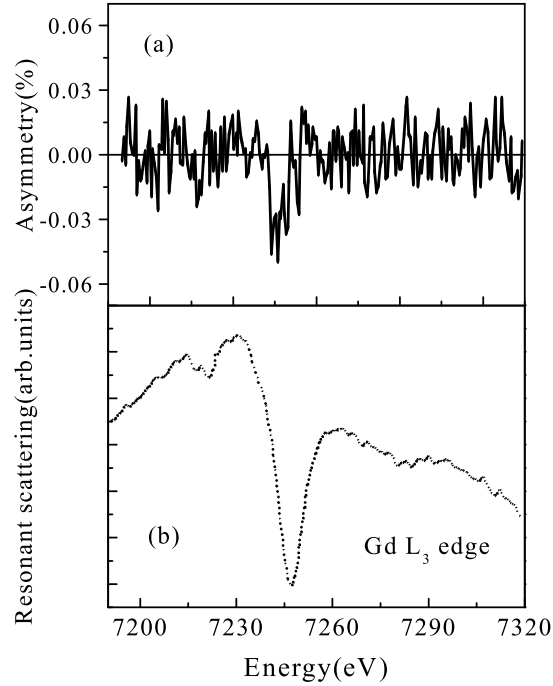}\end{center}

\caption{Intensity asymmetry ratio (a) and X-ray resonant scattering (b) at
the $Gd$ $L_{3}$ edge for the multilayer structure $(Co_{1nm}/Gd_{0.2nm})_{40}$
over $SiO_{2}$. The beam of about 80\% circularly polarized photons
reaches the sample with a grazing angle of 0.38 deg.}
\end{figure}

\begin{figure}
\begin{center}\includegraphics{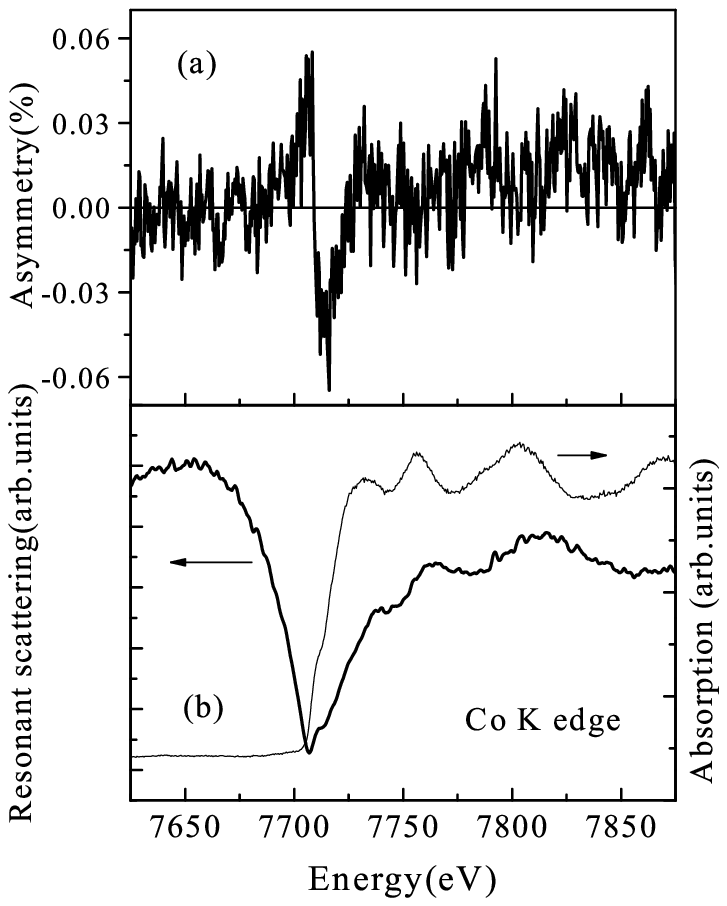}\end{center}

\caption{Intensity asymmetry ratio (a) and X-ray resonant scattering (b) at
the $Co$ $K$ edge for the multilayer structure $(Co_{1nm}/Gd_{0.2nm})_{40}$over
$SiO_{2}$. The beam of about 70\% circularly polarized photons reaches
the sample with a grazing angle of 0.49 deg. The absorption cross
section for $Co$ metallic foil in transmission mode is also shown
in (b).}
\end{figure}

A prototype experiment to study resonant scattering over a wide energy
range has been recently set-up at the dispersive XAS beam line of
LNLS (D06A-DXAS) \cite{Tolentino-PS04}. This beam line is equipped
with an 800 mm-long Rh coated mirror, working at a grazing angle of
4 mrad. The mirror bending mechanism allows vertical beam collimation
or focusing to less than 500 $\mu m$ at the sample position. A curved
$Si(111)$ crystal monochromator selects radiation from a bending
magnet source in the X-ray range from 5 keV up to 14 keV and focuses
it to a 200 $\mu m$ image. The detection is handled with a modified
CCD camera. The energy-direction correlation coming out from the monochromator
is transformed into an energy-position correlation along the lateral
dimension of the CCD detector. Further descriptions of the beam line
are published elsewhere, together with the commissioning results \cite{Tolentino-DXAS04,Tolentino-PS04}. 

Experiments have been performed at grazing incidence to the surface
of the films, close to the external total reflection condition \cite{Nielsen-01}.
In this reflectivity mode, the vertical dimension of the CCD detector
is used to collect the X-ray scattered at different angles. The sample
holder is fit with a goniometer with a horizontal axis, allowing alignment
of the sample at grazing incidence angles with a resolution of 0.005
degrees. A 100 $\mu m$ slit in front of the sample holder limited
the vertical beam. An electromagnet was used to apply a field of up
to 500 $G$ in the plane of the sample and along the propagation direction
of the beam. The asymmetry ratio $R_{XRMS}$ is obtained by flipping
the applied field. 

The preliminary experiment was performed on a multilayer structure
formed by the alternate deposition of 1 nm of $Co$ and 0.2 nm of
$Gd$, repeated 40 times. Both the $Co$ $K$ and the $Gd$ $L_{3}$
edges are accessible and close to each other (7709 eV and 7243 eV),
keeping very similar experimental conditions. $Co$ is present in
large amounts (about 40 nm) and carries a high magnetic moment while
the amount of $Gd$ is reduced (about 30 atomic layers) and carries
a low magnetic moment at room temperature. On the other hand, the
dichroism effect on the $Co$ $K$ edge is much smaller than for $Gd$
$L_{3}$ edge, so that similar signal-over-noise ratios are expected
at the two edges. 

The X-ray resonant scattering intensity around the $Gd$ $L_{3}$
edge was measured by scanning the grazing angle at the sample and
keeping the detector fixed. The critical angle for total external
reflection as well as the Kiessig fringes related to the film thickness
can be clearly identified over the whole energy range (Fig.6). This
result matches very well the calculations based on the multilayer
structure. The total collection time of the two-dimensional figure
was around 15 minutes. 

For the magnetic measurements at the $Gd$ $L_{3}$ edge, the circularly
polarized radiation was selected at 0.18 mrad $(\gamma\Psi\simeq0.49)$
above the orbit plane. The intensity, compared to the maximum at the
orbit plane, was reduced by a factor of three and an elliptically
polarized beam with about 80\% of circular polarization rate was expected
(Fig.4). The grazing incidence angle was set close to the critical
angle of reflection at $\alpha\simeq0.38$ degrees. The superior stability
of the dispersive set-up enables the collection of dichroic spectra
over a wide energy range with a noise smaller than 0.03\%. An asymmetry
ratio of about 0.05\% is identified after a few minutes of acquisition
(Fig.7). The total collection time of this spectrum was less than
10 minutes, with four spectra of 100 s each.

At the $Co$ $K$ edge the circularly polarized radiation was selected
at 0.14 mrad $(\gamma\Psi\simeq0.38)$ above the orbit plane. The
intensity was reduced by a factor of two compared to the maximum at
$\Psi=0$ and an elliptically polarized beam with about 70\% of circular
polarization rate was expected. The grazing incidence angle was set
close to the critical angle of reflection at $\alpha\simeq0.49$ degrees.
The noise level was about 0.01\% after collection of 128 spectra of
30 s each. An asymmetry ratio of about 0.5\% is identified after a
one hour of acquisition (Fig.8). 

We should emphasize that XRMS experiments using dispersive XAS set-up
combine two major advantages : the versatility of sample environment
due to the simple atmospheric working condition in the hard X-ray
range and absence of movements during acquisition characteristic of
dispersive setup.

\section{Concluding Remarks}

We have given here an overview of the prominent aspects of the polarization
of the light delivered by a bending magnet, and some dichroic properties
in X-ray absorption spectroscopy. We presented two applications using
the two XAS beamlines at LNLS. The first example was a combination
of experimental measurements and \emph{ab initio} calculations of
angle resolved X-ray absorption, using linear polarized light to investigate
the local scale structural distortion induced by substrate strain
in $La_{0.7}Sr_{0.3}MnO_{6}$ films. The analysis of the dichroic
experimental data enables the derivation of a model of local distortion
of the coordination shell around the manganese atoms, without any
modification of the tilt angle $Mn-O-Mn$. In the second example we
shown the feasibility of XRMS experiments in thin multilayers using
the dispersive XAS set-up, to get information on local magnetic moments
of rare-earths and transition metal oxides, using the circularly polarized
light delivered by a bending magnet. These preliminary experiments
open the way for more complex studies involving in-situ conditions,
like growth processes or thermal annealing, taking advantage of the
versatility of sample environment and the remarkable experimental
stability.

\subsection*{Acknowledgements}

\begin{acknowledgments}
We are very grateful to Maurizio Sacchi for providing the Gd/Co samples
and for his invaluable support during the first XRMS experiments.
We also acknowledge the LNLS technical staff, especially E. Tamura
and R. Neueschwander for the help in setting up the DXAS experiment,
and F. C. Vicentin for his assistance in the calculations of light
polarization. This work is partially supported by LNLS/ABTLuS/MCT
and FAPESP (1999/12330-6). JCC, NMSN and AYR acknowledge respectively
the grants from FAPESP, CAPES and CNPq. 
\end{acknowledgments}
\bibliographystyle{apsrev}

\begin{thebibliography}{48}
\expandafter\ifx\csname natexlab\endcsname\relax\def\natexlab#1{#1}\fi
\expandafter\ifx\csname bibnamefont\endcsname\relax
  \def\bibnamefont#1{#1}\fi
\expandafter\ifx\csname bibfnamefont\endcsname\relax
  \def\bibfnamefont#1{#1}\fi
\expandafter\ifx\csname citenamefont\endcsname\relax
  \def\citenamefont#1{#1}\fi
\expandafter\ifx\csname url\endcsname\relax
  \def\url#1{\texttt{#1}}\fi
\expandafter\ifx\csname urlprefix\endcsname\relax\def\urlprefix{URL }\fi
\providecommand{\bibinfo}[2]{#2}
\providecommand{\eprint}[2][]{\url{#2}}

\bibitem[{\citenamefont{Jackson}(1975)}]{Jackson-75}
\bibinfo{author}{\bibfnamefont{J.~D.} \bibnamefont{Jackson}},
  \emph{\bibinfo{title}{Classical Electrodynamics, chap. 14}}
  (\bibinfo{publisher}{John Wiley and Sons, New York}, \bibinfo{year}{1975}).

\bibitem[{\citenamefont{Raoux}(1993)}]{Raoux-Hercules93}
\bibinfo{author}{\bibfnamefont{D.}~\bibnamefont{Raoux}},
  \emph{\bibinfo{title}{Introduction to synchrotron radiation and to the
  physics of storage rings}} (\bibinfo{publisher}{Les Editions de Physique,
  Springer Verlag}, \bibinfo{year}{1993}), vol.~\bibinfo{volume}{I},
  chap.~\bibinfo{chapter}{II}, pp. \bibinfo{pages}{37--78}.

\bibitem[{\citenamefont{Rodrigues et~al.}(1998)\citenamefont{Rodrigues,
  Craievich, and da~Silva}}]{Rodrigues-JSR98}
\bibinfo{author}{\bibfnamefont{A.~R.~D.} \bibnamefont{Rodrigues}},
  \bibinfo{author}{\bibfnamefont{A.~F.} \bibnamefont{Craievich}},
  \bibnamefont{and} \bibinfo{author}{\bibfnamefont{C.~E. T.~G.}
  \bibnamefont{da~Silva}}, \bibinfo{journal}{J. Synchr. Radiation}
  \textbf{\bibinfo{volume}{5}}, \bibinfo{pages}{1157} (\bibinfo{year}{1998}).

\bibitem[{\citenamefont{Als-Nielsen and McMorrow}(2001)}]{Nielsen-01}
\bibinfo{author}{\bibfnamefont{J.}~\bibnamefont{Als-Nielsen}} \bibnamefont{and}
  \bibinfo{author}{\bibfnamefont{D.}~\bibnamefont{McMorrow}},
  \emph{\bibinfo{title}{Refraction and reflection from interfaces}}
  (\bibinfo{publisher}{John Wiley and Sons}, \bibinfo{year}{2001}),
  chap.~\bibinfo{chapter}{3}, pp. \bibinfo{pages}{61--106}.

\bibitem[{\citenamefont{Giles et~al.}(1994)\citenamefont{Giles, Malgrange,
  Goulon, DeBergevin, Vettier, Dartyge, Fontaine, Giogetti, and
  Pizzini}}]{Giles-JAC94}
\bibinfo{author}{\bibfnamefont{C.}~\bibnamefont{Giles}},
  \bibinfo{author}{\bibfnamefont{C.}~\bibnamefont{Malgrange}},
  \bibinfo{author}{\bibfnamefont{J.}~\bibnamefont{Goulon}},
  \bibinfo{author}{\bibfnamefont{F.}~\bibnamefont{DeBergevin}},
  \bibinfo{author}{\bibfnamefont{C.}~\bibnamefont{Vettier}},
  \bibinfo{author}{\bibfnamefont{E.}~\bibnamefont{Dartyge}},
  \bibinfo{author}{\bibfnamefont{A.}~\bibnamefont{Fontaine}},
  \bibinfo{author}{\bibfnamefont{C.}~\bibnamefont{Giogetti}}, \bibnamefont{and}
  \bibinfo{author}{\bibfnamefont{S.}~\bibnamefont{Pizzini}},
  \bibinfo{journal}{J. Appl. Cryst.} \textbf{\bibinfo{volume}{27}},
  \bibinfo{pages}{232} (\bibinfo{year}{1994}).

\bibitem[{\citenamefont{James}(1965)}]{James-65}
\bibinfo{author}{\bibfnamefont{R.~W.} \bibnamefont{James}},
  \emph{\bibinfo{title}{The optical principles of the diffraction of x-rays}}
  (\bibinfo{publisher}{G. Bell and sons LTD, London UK}, \bibinfo{year}{1965}).

\bibitem[{\citenamefont{Dartyge
  et~al.}(1986{\natexlab{a}})\citenamefont{Dartyge, Depautex, Dubuisson,
  Fontaine, Jucha, Leboucher, and Tourillon}}]{Dartyge-NIM86}
\bibinfo{author}{\bibfnamefont{E.}~\bibnamefont{Dartyge}},
  \bibinfo{author}{\bibfnamefont{C.}~\bibnamefont{Depautex}},
  \bibinfo{author}{\bibfnamefont{J.}~\bibnamefont{Dubuisson}},
  \bibinfo{author}{\bibfnamefont{A.}~\bibnamefont{Fontaine}},
  \bibinfo{author}{\bibfnamefont{A.}~\bibnamefont{Jucha}},
  \bibinfo{author}{\bibfnamefont{P.}~\bibnamefont{Leboucher}},
  \bibnamefont{and}
  \bibinfo{author}{\bibfnamefont{G.}~\bibnamefont{Tourillon}},
  \bibinfo{journal}{Nucl. Instrum. Methods A} \textbf{\bibinfo{volume}{246}},
  \bibinfo{pages}{452} (\bibinfo{year}{1986}{\natexlab{a}}).

\bibitem[{\citenamefont{Koningsberger and Prins}(1988)}]{Koningsberger-88}
\bibinfo{editor}{\bibfnamefont{D.~C.} \bibnamefont{Koningsberger}}
  \bibnamefont{and} \bibinfo{editor}{\bibfnamefont{R.}~\bibnamefont{Prins}},
  eds. (\bibinfo{publisher}{Wiley-Intersciences, New York},
  \bibinfo{year}{1988}).

\bibitem[{\citenamefont{Fontaine}(1993)}]{Fontaine-Hercules93}
\bibinfo{author}{\bibfnamefont{A.}~\bibnamefont{Fontaine}},
  \emph{\bibinfo{title}{Interaction of X-Rays with Matter: X-Ray Absorption
  Spectroscopy}} (\bibinfo{publisher}{Les Editions de Physique, Springer
  Verlag}, \bibinfo{year}{1993}), vol.~\bibinfo{volume}{I},
  chap.~\bibinfo{chapter}{XV}, pp. \bibinfo{pages}{323--369}.

\bibitem[{\citenamefont{Natoli et~al.}(1980)\citenamefont{Natoli, Misemer,
  Doniach, and Kutzler}}]{Natoli-PRA80}
\bibinfo{author}{\bibfnamefont{C.~R.} \bibnamefont{Natoli}},
  \bibinfo{author}{\bibfnamefont{D.~K.} \bibnamefont{Misemer}},
  \bibinfo{author}{\bibfnamefont{S.}~\bibnamefont{Doniach}}, \bibnamefont{and}
  \bibinfo{author}{\bibfnamefont{F.~W.} \bibnamefont{Kutzler}},
  \bibinfo{journal}{Phys. Rev. A} \textbf{\bibinfo{volume}{22}},
  \bibinfo{pages}{1104} (\bibinfo{year}{1980}).

\bibitem[{\citenamefont{Rehr and Albers}(2000)}]{Rehr-RMP00}
\bibinfo{author}{\bibfnamefont{J.~J.} \bibnamefont{Rehr}} \bibnamefont{and}
  \bibinfo{author}{\bibfnamefont{R.~C.} \bibnamefont{Albers}},
  \bibinfo{journal}{Rev. Mod. Phys.} \textbf{\bibinfo{volume}{72}},
  \bibinfo{pages}{621} (\bibinfo{year}{2000}).

\bibitem[{\citenamefont{DeGroot}(1994)}]{DeGroot-JESRP94}
\bibinfo{author}{\bibfnamefont{F.~M.~F.} \bibnamefont{DeGroot}},
  \bibinfo{journal}{Journal Elect. Spect. Relat. Phenom.}
  \textbf{\bibinfo{volume}{67}}, \bibinfo{pages}{529} (\bibinfo{year}{1994}).

\bibitem[{\citenamefont{Brouder}(1990)}]{Brouder-JPCM90}
\bibinfo{author}{\bibfnamefont{C.}~\bibnamefont{Brouder}}, \bibinfo{journal}{J.
  Phys. Condens. Matter} \textbf{\bibinfo{volume}{2}}, \bibinfo{pages}{701}
  (\bibinfo{year}{1990}).

\bibitem[{\citenamefont{Stern}(1974)}]{Stern-PRB74}
\bibinfo{author}{\bibfnamefont{E.~A.} \bibnamefont{Stern}},
  \bibinfo{journal}{Phys. Rev. B} \textbf{\bibinfo{volume}{10}},
  \bibinfo{pages}{3027} (\bibinfo{year}{1974}).

\bibitem[{\citenamefont{Heald and Stern}(1977)}]{Heald-77}
\bibinfo{author}{\bibfnamefont{S.~M.} \bibnamefont{Heald}} \bibnamefont{and}
  \bibinfo{author}{\bibfnamefont{E.~A.} \bibnamefont{Stern}},
  \bibinfo{journal}{Phys. Rev. B} \textbf{\bibinfo{volume}{46}},
  \bibinfo{pages}{5549} (\bibinfo{year}{1977}).

\bibitem[{\citenamefont{Haskel}(1998)}]{Haskel-Phd98}
\bibinfo{author}{\bibfnamefont{D.}~\bibnamefont{Haskel}}, Ph.D. thesis,
  \bibinfo{school}{University of Washington} (\bibinfo{year}{1998}).

\bibitem[{\citenamefont{Magnan et~al.}(1991)\citenamefont{Magnan, Chandesris,
  Villette, Heckmann, and Lecante}}]{Magnan-PRL91}
\bibinfo{author}{\bibfnamefont{H.}~\bibnamefont{Magnan}},
  \bibinfo{author}{\bibfnamefont{D.}~\bibnamefont{Chandesris}},
  \bibinfo{author}{\bibfnamefont{B.}~\bibnamefont{Villette}},
  \bibinfo{author}{\bibfnamefont{O.}~\bibnamefont{Heckmann}}, \bibnamefont{and}
  \bibinfo{author}{\bibfnamefont{J.}~\bibnamefont{Lecante}},
  \bibinfo{journal}{Phys. Rev. Lett} \textbf{\bibinfo{volume}{67}},
  \bibinfo{pages}{259} (\bibinfo{year}{1991}).

\bibitem[{\citenamefont{Tolentino et~al.}(1992)\citenamefont{Tolentino,
  Baudelet, Fontaine, Gourieux, Krill, Henry, and
  Rossat-Mignod}}]{Tolentino-PhyC92}
\bibinfo{author}{\bibfnamefont{H.}~\bibnamefont{Tolentino}},
  \bibinfo{author}{\bibfnamefont{F.}~\bibnamefont{Baudelet}},
  \bibinfo{author}{\bibfnamefont{A.}~\bibnamefont{Fontaine}},
  \bibinfo{author}{\bibfnamefont{T.}~\bibnamefont{Gourieux}},
  \bibinfo{author}{\bibfnamefont{G.}~\bibnamefont{Krill}},
  \bibinfo{author}{\bibfnamefont{J.~Y.} \bibnamefont{Henry}}, \bibnamefont{and}
  \bibinfo{author}{\bibfnamefont{J.}~\bibnamefont{Rossat-Mignod}},
  \bibinfo{journal}{Physica C} \textbf{\bibinfo{volume}{192}},
  \bibinfo{pages}{115} (\bibinfo{year}{1992}).

\bibitem[{\citenamefont{Gaudry et~al.}(2003)\citenamefont{Gaudry, Kiratisin,
  Sainctavit, Brouder, Mauri, Ramos, Rogalev, and Goulon}}]{Gaudry-PRB03}
\bibinfo{author}{\bibfnamefont{E.}~\bibnamefont{Gaudry}},
  \bibinfo{author}{\bibfnamefont{A.}~\bibnamefont{Kiratisin}},
  \bibinfo{author}{\bibfnamefont{P.}~\bibnamefont{Sainctavit}},
  \bibinfo{author}{\bibfnamefont{C.}~\bibnamefont{Brouder}},
  \bibinfo{author}{\bibfnamefont{F.}~\bibnamefont{Mauri}},
  \bibinfo{author}{\bibfnamefont{A.}~\bibnamefont{Ramos}},
  \bibinfo{author}{\bibfnamefont{A.}~\bibnamefont{Rogalev}}, \bibnamefont{and}
  \bibinfo{author}{\bibfnamefont{J.}~\bibnamefont{Goulon}},
  \bibinfo{journal}{Phys. Rev. B} \textbf{\bibinfo{volume}{67}},
  \bibinfo{pages}{094108} (\bibinfo{year}{2003}).

\bibitem[{\citenamefont{Pizzini et~al.}(1992)\citenamefont{Pizzini, Baudelet,
  Chandesris, Fontaine, Magnan, George, Petroff, Barthelemy, Fert, Loloee
  et~al.}}]{Pizzini-PRBRC92}
\bibinfo{author}{\bibfnamefont{S.}~\bibnamefont{Pizzini}},
  \bibinfo{author}{\bibfnamefont{F.}~\bibnamefont{Baudelet}},
  \bibinfo{author}{\bibfnamefont{D.}~\bibnamefont{Chandesris}},
  \bibinfo{author}{\bibfnamefont{A.}~\bibnamefont{Fontaine}},
  \bibinfo{author}{\bibfnamefont{H.}~\bibnamefont{Magnan}},
  \bibinfo{author}{\bibfnamefont{J.~M.} \bibnamefont{George}},
  \bibinfo{author}{\bibfnamefont{F.}~\bibnamefont{Petroff}},
  \bibinfo{author}{\bibfnamefont{A.}~\bibnamefont{Barthelemy}},
  \bibinfo{author}{\bibfnamefont{A.}~\bibnamefont{Fert}},
  \bibinfo{author}{\bibfnamefont{R.}~\bibnamefont{Loloee}},
  \bibnamefont{et~al.}, \bibinfo{journal}{Phys. Rev. B}
  \textbf{\bibinfo{volume}{46}}, \bibinfo{pages}{1253} (\bibinfo{year}{1992}).

\bibitem[{\citenamefont{Souza-Neto et~al.}(2003)\citenamefont{Souza-Neto,
  Ramos, Tolentino, Favre-Nicolin, and Ranno}}]{Souza-Neto-APL03}
\bibinfo{author}{\bibfnamefont{N.~M.} \bibnamefont{Souza-Neto}},
  \bibinfo{author}{\bibfnamefont{A.~Y.} \bibnamefont{Ramos}},
  \bibinfo{author}{\bibfnamefont{H.~C.~N.} \bibnamefont{Tolentino}},
  \bibinfo{author}{\bibfnamefont{E.}~\bibnamefont{Favre-Nicolin}},
  \bibnamefont{and} \bibinfo{author}{\bibfnamefont{L.}~\bibnamefont{Ranno}},
  \bibinfo{journal}{Appl. Phys. Lett.} \textbf{\bibinfo{volume}{83}},
  \bibinfo{pages}{3587} (\bibinfo{year}{2003}).

\bibitem[{\citenamefont{Ebert}(1996)}]{Ebert-96}
\bibinfo{author}{\bibfnamefont{H.}~\bibnamefont{Ebert}},
  \emph{\bibinfo{title}{Circular magnetic X-ray dichroism in transition metal
  systems}} (\bibinfo{publisher}{Springer, Berlin}, \bibinfo{year}{1996}), pp.
  \bibinfo{pages}{159--177}.

\bibitem[{\citenamefont{Schütz et~al.}(1987)\citenamefont{Schütz, Wagner,
  Wilhelm, and Kienle}}]{Schutz-87}
\bibinfo{author}{\bibfnamefont{G.}~\bibnamefont{Schütz}},
  \bibinfo{author}{\bibfnamefont{W.}~\bibnamefont{Wagner}},
  \bibinfo{author}{\bibfnamefont{W.}~\bibnamefont{Wilhelm}}, \bibnamefont{and}
  \bibinfo{author}{\bibfnamefont{P.}~\bibnamefont{Kienle}},
  \bibinfo{journal}{Phys. Rev. Lett.} \textbf{\bibinfo{volume}{58}},
  \bibinfo{pages}{737} (\bibinfo{year}{1987}).

\bibitem[{\citenamefont{Baudelet et~al.}(1991)\citenamefont{Baudelet, Dartyge,
  Fontaine, Brouder, Krill, Kapller, and Piecuch}}]{Baudelet-PRB91}
\bibinfo{author}{\bibfnamefont{F.}~\bibnamefont{Baudelet}},
  \bibinfo{author}{\bibfnamefont{E.}~\bibnamefont{Dartyge}},
  \bibinfo{author}{\bibfnamefont{A.}~\bibnamefont{Fontaine}},
  \bibinfo{author}{\bibfnamefont{C.}~\bibnamefont{Brouder}},
  \bibinfo{author}{\bibfnamefont{G.}~\bibnamefont{Krill}},
  \bibinfo{author}{\bibfnamefont{J.~P.} \bibnamefont{Kapller}},
  \bibnamefont{and} \bibinfo{author}{\bibfnamefont{M.}~\bibnamefont{Piecuch}},
  \bibinfo{journal}{Phys. Rev. B} \textbf{\bibinfo{volume}{43}},
  \bibinfo{pages}{5857} (\bibinfo{year}{1991}).

\bibitem[{\citenamefont{Giorgetti et~al.}(1993)\citenamefont{Giorgetti,
  Pizzini, Dartyge, Fontaine, Baudelet, Brouder, Bauer, Krill, Miraglia,
  Fruchart et~al.}}]{Giorgetti-PRB93}
\bibinfo{author}{\bibfnamefont{C.}~\bibnamefont{Giorgetti}},
  \bibinfo{author}{\bibfnamefont{S.}~\bibnamefont{Pizzini}},
  \bibinfo{author}{\bibfnamefont{E.}~\bibnamefont{Dartyge}},
  \bibinfo{author}{\bibfnamefont{A.}~\bibnamefont{Fontaine}},
  \bibinfo{author}{\bibfnamefont{F.}~\bibnamefont{Baudelet}},
  \bibinfo{author}{\bibfnamefont{C.}~\bibnamefont{Brouder}},
  \bibinfo{author}{\bibfnamefont{P.}~\bibnamefont{Bauer}},
  \bibinfo{author}{\bibfnamefont{G.}~\bibnamefont{Krill}},
  \bibinfo{author}{\bibfnamefont{S.}~\bibnamefont{Miraglia}},
  \bibinfo{author}{\bibfnamefont{D.}~\bibnamefont{Fruchart}},
  \bibnamefont{et~al.}, \bibinfo{journal}{Phys. Rev. B}
  \textbf{\bibinfo{volume}{48}}, \bibinfo{pages}{12732} (\bibinfo{year}{1993}).

\bibitem[{\citenamefont{Gibbs et~al.}(1988)\citenamefont{Gibbs, Harshman,
  Isaacs, McWhan, Mills, and Vettier}}]{Gibbs-PRL88}
\bibinfo{author}{\bibfnamefont{D.}~\bibnamefont{Gibbs}},
  \bibinfo{author}{\bibfnamefont{D.~R.} \bibnamefont{Harshman}},
  \bibinfo{author}{\bibfnamefont{E.}~\bibnamefont{Isaacs}},
  \bibinfo{author}{\bibfnamefont{D.~B.} \bibnamefont{McWhan}},
  \bibinfo{author}{\bibfnamefont{D.}~\bibnamefont{Mills}}, \bibnamefont{and}
  \bibinfo{author}{\bibfnamefont{C.}~\bibnamefont{Vettier}},
  \bibinfo{journal}{Phys. Rev. Lett.} \textbf{\bibinfo{volume}{61}},
  \bibinfo{pages}{1241} (\bibinfo{year}{1988}).

\bibitem[{\citenamefont{Hannon et~al.}(1988)\citenamefont{Hannon, Trammell,
  Blume, and Gibbs}}]{Hannon-PRL88}
\bibinfo{author}{\bibfnamefont{J.~P.} \bibnamefont{Hannon}},
  \bibinfo{author}{\bibfnamefont{G.~T.} \bibnamefont{Trammell}},
  \bibinfo{author}{\bibfnamefont{M.}~\bibnamefont{Blume}}, \bibnamefont{and}
  \bibinfo{author}{\bibfnamefont{D.}~\bibnamefont{Gibbs}},
  \bibinfo{journal}{Phys. Rev. Lett.} \textbf{\bibinfo{volume}{61}},
  \bibinfo{pages}{1245} (\bibinfo{year}{1988}).

\bibitem[{\citenamefont{Fontcuberta et~al.}(1996)\citenamefont{Fontcuberta,
  Martinez, Seffar, Pinol, Garcia-Munoz, and Obradors}}]{Fontcuberta-PRL96}
\bibinfo{author}{\bibfnamefont{J.}~\bibnamefont{Fontcuberta}},
  \bibinfo{author}{\bibfnamefont{B.}~\bibnamefont{Martinez}},
  \bibinfo{author}{\bibfnamefont{A.}~\bibnamefont{Seffar}},
  \bibinfo{author}{\bibfnamefont{S.}~\bibnamefont{Pinol}},
  \bibinfo{author}{\bibfnamefont{J.}~\bibnamefont{Garcia-Munoz}},
  \bibnamefont{and} \bibinfo{author}{\bibfnamefont{X.}~\bibnamefont{Obradors}},
  \bibinfo{journal}{Phys. Rev. Lett} \textbf{\bibinfo{volume}{76}},
  \bibinfo{pages}{1122} (\bibinfo{year}{1996}).

\bibitem[{\citenamefont{Millis}(1998)}]{Millis-N98}
\bibinfo{author}{\bibfnamefont{A.~J.} \bibnamefont{Millis}},
  \bibinfo{journal}{Nature} \textbf{\bibinfo{volume}{392}},
  \bibinfo{pages}{147} (\bibinfo{year}{1998}).

\bibitem[{\citenamefont{Salamon and Jaime}(2001)}]{Salamon-RMP01}
\bibinfo{author}{\bibfnamefont{M.~B.} \bibnamefont{Salamon}} \bibnamefont{and}
  \bibinfo{author}{\bibfnamefont{M.}~\bibnamefont{Jaime}},
  \bibinfo{journal}{Rev. Mod. Phys.} \textbf{\bibinfo{volume}{73}},
  \bibinfo{pages}{583} (\bibinfo{year}{2001}).

\bibitem[{\citenamefont{Millis et~al.}(1998)\citenamefont{Millis, Darling, and
  Migliori}}]{Millis-JAP98}
\bibinfo{author}{\bibfnamefont{A.~J.} \bibnamefont{Millis}},
  \bibinfo{author}{\bibfnamefont{T.}~\bibnamefont{Darling}}, \bibnamefont{and}
  \bibinfo{author}{\bibfnamefont{A.}~\bibnamefont{Migliori}},
  \bibinfo{journal}{J. Appl. Phys.} \textbf{\bibinfo{volume}{83}},
  \bibinfo{pages}{1588} (\bibinfo{year}{1998}).

\bibitem[{\citenamefont{Prellier et~al.}(2001)\citenamefont{Prellier, Lecoeur,
  and Mercey}}]{Prellier-JPCM01}
\bibinfo{author}{\bibfnamefont{W.}~\bibnamefont{Prellier}},
  \bibinfo{author}{\bibfnamefont{P.}~\bibnamefont{Lecoeur}}, \bibnamefont{and}
  \bibinfo{author}{\bibfnamefont{B.}~\bibnamefont{Mercey}},
  \bibinfo{journal}{J. Phys.: Condens. Matter} \textbf{\bibinfo{volume}{13}},
  \bibinfo{pages}{R915} (\bibinfo{year}{2001}).

\bibitem[{\citenamefont{Ranno et~al.}(2002)\citenamefont{Ranno, Llobet, Tiron,
  and Favre-Nicolin}}]{Ranno-ASS02}
\bibinfo{author}{\bibfnamefont{L.}~\bibnamefont{Ranno}},
  \bibinfo{author}{\bibfnamefont{A.}~\bibnamefont{Llobet}},
  \bibinfo{author}{\bibfnamefont{R.}~\bibnamefont{Tiron}}, \bibnamefont{and}
  \bibinfo{author}{\bibfnamefont{E.}~\bibnamefont{Favre-Nicolin}},
  \bibinfo{journal}{Applied Surface Science} \textbf{\bibinfo{volume}{188}},
  \bibinfo{pages}{170} (\bibinfo{year}{2002}).

\bibitem[{\citenamefont{Natoli}(1984)}]{Natoli-84}
\bibinfo{author}{\bibfnamefont{C.}~\bibnamefont{Natoli}},
  \emph{\bibinfo{title}{Distance dependence of continuum and bound state of
  excitonic resonance in X-ray absorption near edge structure (XANES)}}
  (\bibinfo{publisher}{Springer-Verlag}, \bibinfo{year}{1984}),
  chap.~\bibinfo{chapter}{4}, pp. \bibinfo{pages}{38--42}.

\bibitem[{\citenamefont{Tolentino et~al.}(2001)\citenamefont{Tolentino, Ramos,
  Alves, Barrea, Tamura, Cezar, and Watanabe}}]{Tolentino-JSR01}
\bibinfo{author}{\bibfnamefont{H.~C.~N.} \bibnamefont{Tolentino}},
  \bibinfo{author}{\bibfnamefont{A.~Y.} \bibnamefont{Ramos}},
  \bibinfo{author}{\bibfnamefont{M.~C.~M.} \bibnamefont{Alves}},
  \bibinfo{author}{\bibfnamefont{R.~A.} \bibnamefont{Barrea}},
  \bibinfo{author}{\bibfnamefont{E.}~\bibnamefont{Tamura}},
  \bibinfo{author}{\bibfnamefont{J.~C.} \bibnamefont{Cezar}}, \bibnamefont{and}
  \bibinfo{author}{\bibfnamefont{N.}~\bibnamefont{Watanabe}},
  \bibinfo{journal}{J. Synchrotron Rad.} \textbf{\bibinfo{volume}{8}},
  \bibinfo{pages}{1040} (\bibinfo{year}{2001}).

\bibitem[{\citenamefont{Souza-Neto
  et~al.}(2004{\natexlab{a}})\citenamefont{Souza-Neto, Ramos, Tolentino,
  Favre-Nicolin, and Ranno}}]{SouzaNeto-PRB04}
\bibinfo{author}{\bibfnamefont{N.~M.} \bibnamefont{Souza-Neto}},
  \bibinfo{author}{\bibfnamefont{A.~Y.} \bibnamefont{Ramos}},
  \bibinfo{author}{\bibfnamefont{H.~C.~N.} \bibnamefont{Tolentino}},
  \bibinfo{author}{\bibfnamefont{E.}~\bibnamefont{Favre-Nicolin}},
  \bibnamefont{and} \bibinfo{author}{\bibfnamefont{L.}~\bibnamefont{Ranno}},
  \bibinfo{journal}{Phys. Rev. B}  (\bibinfo{year}{2004}{\natexlab{a}}),
  \bibinfo{note}{in press}.

\bibitem[{\citenamefont{Ankudinov et~al.}(2002)\citenamefont{Ankudinov,
  Bouldin, Rehr, Sims, and Hung}}]{Ankudinov-PRB02}
\bibinfo{author}{\bibfnamefont{A.~L.} \bibnamefont{Ankudinov}},
  \bibinfo{author}{\bibfnamefont{C.}~\bibnamefont{Bouldin}},
  \bibinfo{author}{\bibfnamefont{J.~J.} \bibnamefont{Rehr}},
  \bibinfo{author}{\bibfnamefont{J.}~\bibnamefont{Sims}}, \bibnamefont{and}
  \bibinfo{author}{\bibfnamefont{H.}~\bibnamefont{Hung}},
  \bibinfo{journal}{Phys. Rev. B} \textbf{\bibinfo{volume}{65}},
  \bibinfo{pages}{104107} (\bibinfo{year}{2002}).

\bibitem[{\citenamefont{Souza-Neto
  et~al.}(2004{\natexlab{b}})\citenamefont{Souza-Neto, Ramos, Tolentino,
  Favre-Nicolin, and Ranno}}]{Souza-Neto-JAllC04}
\bibinfo{author}{\bibfnamefont{N.~M.} \bibnamefont{Souza-Neto}},
  \bibinfo{author}{\bibfnamefont{A.~Y.} \bibnamefont{Ramos}},
  \bibinfo{author}{\bibfnamefont{H.~C.~N.} \bibnamefont{Tolentino}},
  \bibinfo{author}{\bibfnamefont{E.}~\bibnamefont{Favre-Nicolin}},
  \bibnamefont{and} \bibinfo{author}{\bibfnamefont{L.}~\bibnamefont{Ranno}},
  \bibinfo{journal}{J. of Alloy and Comp.} \textbf{\bibinfo{volume}{369}},
  \bibinfo{pages}{205} (\bibinfo{year}{2004}{\natexlab{b}}).

\bibitem[{\citenamefont{Souza-Neto
  et~al.}(2004{\natexlab{c}})\citenamefont{Souza-Neto, Ramos, Tolentino,
  Favre-Nicolin, and Ranno}}]{SouzaNeto-PS04}
\bibinfo{author}{\bibfnamefont{N.~M.} \bibnamefont{Souza-Neto}},
  \bibinfo{author}{\bibfnamefont{A.~Y.} \bibnamefont{Ramos}},
  \bibinfo{author}{\bibfnamefont{H.~C.~N.} \bibnamefont{Tolentino}},
  \bibinfo{author}{\bibfnamefont{E.}~\bibnamefont{Favre-Nicolin}},
  \bibnamefont{and} \bibinfo{author}{\bibfnamefont{L.}~\bibnamefont{Ranno}},
  \bibinfo{journal}{Physica Scripta}  (\bibinfo{year}{2004}{\natexlab{c}}),
  \bibinfo{note}{in press}.

\bibitem[{\citenamefont{Ramos et~al.}(2003)\citenamefont{Ramos, Souza-Neto,
  Giacomelli, Tolentino, Ranno, and Favre-Nicolin}}]{Ramos-ROMA02}
\bibinfo{author}{\bibfnamefont{A.~Y.} \bibnamefont{Ramos}},
  \bibinfo{author}{\bibfnamefont{N.~M.} \bibnamefont{Souza-Neto}},
  \bibinfo{author}{\bibfnamefont{C.}~\bibnamefont{Giacomelli}},
  \bibinfo{author}{\bibfnamefont{H.~C.~N.} \bibnamefont{Tolentino}},
  \bibinfo{author}{\bibfnamefont{L.}~\bibnamefont{Ranno}}, \bibnamefont{and}
  \bibinfo{author}{\bibfnamefont{E.}~\bibnamefont{Favre-Nicolin}},
  \bibinfo{journal}{AIP Conference Proceedings} \textbf{\bibinfo{volume}{652}},
  \bibinfo{pages}{456} (\bibinfo{year}{2003}).

\bibitem[{\citenamefont{Kao et~al.}(1990)\citenamefont{Kao, Hastings, Johnson,
  Siddons, Smith, and Prinz}}]{Kao-PRL90}
\bibinfo{author}{\bibfnamefont{C.~C.} \bibnamefont{Kao}},
  \bibinfo{author}{\bibfnamefont{J.~B.} \bibnamefont{Hastings}},
  \bibinfo{author}{\bibfnamefont{E.~D.} \bibnamefont{Johnson}},
  \bibinfo{author}{\bibfnamefont{D.~P.} \bibnamefont{Siddons}},
  \bibinfo{author}{\bibfnamefont{G.~C.} \bibnamefont{Smith}}, \bibnamefont{and}
  \bibinfo{author}{\bibfnamefont{G.~A.} \bibnamefont{Prinz}},
  \bibinfo{journal}{Phys. Rev. Lett.} \textbf{\bibinfo{volume}{65}},
  \bibinfo{pages}{373} (\bibinfo{year}{1990}).

\bibitem[{\citenamefont{Tonnerre et~al.}(1995)\citenamefont{Tonnerre, Seve,
  Raoux, Soullie, Rodmacq, and Wolfers}}]{Tonnerre-PRL95}
\bibinfo{author}{\bibfnamefont{J.~M.} \bibnamefont{Tonnerre}},
  \bibinfo{author}{\bibfnamefont{L.}~\bibnamefont{Seve}},
  \bibinfo{author}{\bibfnamefont{D.}~\bibnamefont{Raoux}},
  \bibinfo{author}{\bibfnamefont{G.}~\bibnamefont{Soullie}},
  \bibinfo{author}{\bibfnamefont{B.}~\bibnamefont{Rodmacq}}, \bibnamefont{and}
  \bibinfo{author}{\bibfnamefont{P.}~\bibnamefont{Wolfers}},
  \bibinfo{journal}{Phys. Rev. Lett.} \textbf{\bibinfo{volume}{75}},
  \bibinfo{pages}{740} (\bibinfo{year}{1995}).

\bibitem[{\citenamefont{Sacchi et~al.}(1998)\citenamefont{Sacchi, Hague,
  Pasquali, Mironi, Mariot, Isberg, Gullikson, and Unberwood}}]{Sacchi-PRL98}
\bibinfo{author}{\bibfnamefont{M.}~\bibnamefont{Sacchi}},
  \bibinfo{author}{\bibfnamefont{C.~F.} \bibnamefont{Hague}},
  \bibinfo{author}{\bibfnamefont{L.}~\bibnamefont{Pasquali}},
  \bibinfo{author}{\bibfnamefont{A.}~\bibnamefont{Mironi}},
  \bibinfo{author}{\bibfnamefont{J.~M.} \bibnamefont{Mariot}},
  \bibinfo{author}{\bibfnamefont{P.}~\bibnamefont{Isberg}},
  \bibinfo{author}{\bibfnamefont{E.~M.} \bibnamefont{Gullikson}},
  \bibnamefont{and} \bibinfo{author}{\bibfnamefont{J.~H.}
  \bibnamefont{Unberwood}}, \bibinfo{journal}{Phys. Rev. Lett.}
  \textbf{\bibinfo{volume}{81}}, \bibinfo{pages}{1521} (\bibinfo{year}{1998}).

\bibitem[{\citenamefont{Dartyge
  et~al.}(1986{\natexlab{b}})\citenamefont{Dartyge, Fontaine, Tourillon,
  Cortes, and Jucha}}]{Dartyge-PL86}
\bibinfo{author}{\bibfnamefont{E.}~\bibnamefont{Dartyge}},
  \bibinfo{author}{\bibfnamefont{A.}~\bibnamefont{Fontaine}},
  \bibinfo{author}{\bibfnamefont{G.}~\bibnamefont{Tourillon}},
  \bibinfo{author}{\bibfnamefont{R.}~\bibnamefont{Cortes}}, \bibnamefont{and}
  \bibinfo{author}{\bibfnamefont{A.}~\bibnamefont{Jucha}},
  \bibinfo{journal}{Physics Letters A} \textbf{\bibinfo{volume}{113}},
  \bibinfo{pages}{384} (\bibinfo{year}{1986}{\natexlab{b}}).

\bibitem[{\citenamefont{Seve et~al.}(1999)\citenamefont{Seve, Jaouen, Tonnerre,
  Raoux, Bartolome, Arend, Felsch, Rogalev, Goulon, Gautier
  et~al.}}]{Seve-PRB99}
\bibinfo{author}{\bibfnamefont{L.}~\bibnamefont{Seve}},
  \bibinfo{author}{\bibfnamefont{N.}~\bibnamefont{Jaouen}},
  \bibinfo{author}{\bibfnamefont{J.~M.} \bibnamefont{Tonnerre}},
  \bibinfo{author}{\bibfnamefont{D.}~\bibnamefont{Raoux}},
  \bibinfo{author}{\bibfnamefont{F.}~\bibnamefont{Bartolome}},
  \bibinfo{author}{\bibfnamefont{M.}~\bibnamefont{Arend}},
  \bibinfo{author}{\bibfnamefont{W.}~\bibnamefont{Felsch}},
  \bibinfo{author}{\bibfnamefont{A.}~\bibnamefont{Rogalev}},
  \bibinfo{author}{\bibfnamefont{J.}~\bibnamefont{Goulon}},
  \bibinfo{author}{\bibfnamefont{C.}~\bibnamefont{Gautier}},
  \bibnamefont{et~al.}, \bibinfo{journal}{Phys. Rev. B}
  \textbf{\bibinfo{volume}{60}}, \bibinfo{pages}{9662} (\bibinfo{year}{1999}).

\bibitem[{\citenamefont{Jaouen et~al.}(2002)\citenamefont{Jaouen, Tonnerre,
  Raoux, Bontempi, Ortega, m.~Muenzenberg, Felsch, Rogalev, Durr, Dudzik
  et~al.}}]{Jaouen-PRB02}
\bibinfo{author}{\bibfnamefont{N.}~\bibnamefont{Jaouen}},
  \bibinfo{author}{\bibfnamefont{J.~M.} \bibnamefont{Tonnerre}},
  \bibinfo{author}{\bibfnamefont{D.}~\bibnamefont{Raoux}},
  \bibinfo{author}{\bibfnamefont{E.}~\bibnamefont{Bontempi}},
  \bibinfo{author}{\bibfnamefont{L.}~\bibnamefont{Ortega}},
  \bibinfo{author}{\bibnamefont{m.~Muenzenberg}},
  \bibinfo{author}{\bibfnamefont{W.}~\bibnamefont{Felsch}},
  \bibinfo{author}{\bibfnamefont{A.}~\bibnamefont{Rogalev}},
  \bibinfo{author}{\bibfnamefont{H.~A.} \bibnamefont{Durr}},
  \bibinfo{author}{\bibfnamefont{E.}~\bibnamefont{Dudzik}},
  \bibnamefont{et~al.}, \bibinfo{journal}{Phys. Rev. B}
  \textbf{\bibinfo{volume}{66}}, \bibinfo{pages}{134420/1}
  (\bibinfo{year}{2002}).

\bibitem[{\citenamefont{Tolentino
  et~al.}(2004{\natexlab{a}})\citenamefont{Tolentino, Cezar, Watanabe,
  Piamonteze, Souza-Neto, Tamura, Ramos, and Neueschwander}}]{Tolentino-PS04}
\bibinfo{author}{\bibfnamefont{H.~C.~N.} \bibnamefont{Tolentino}},
  \bibinfo{author}{\bibfnamefont{J.~C.} \bibnamefont{Cezar}},
  \bibinfo{author}{\bibfnamefont{N.}~\bibnamefont{Watanabe}},
  \bibinfo{author}{\bibfnamefont{C.}~\bibnamefont{Piamonteze}},
  \bibinfo{author}{\bibfnamefont{N.~M.} \bibnamefont{Souza-Neto}},
  \bibinfo{author}{\bibfnamefont{E.}~\bibnamefont{Tamura}},
  \bibinfo{author}{\bibfnamefont{A.~Y.} \bibnamefont{Ramos}}, \bibnamefont{and}
  \bibinfo{author}{\bibfnamefont{R.}~\bibnamefont{Neueschwander}},
  \bibinfo{journal}{Physica Scripta}  (\bibinfo{year}{2004}{\natexlab{a}}),
  \bibinfo{note}{in press}.

\bibitem[{\citenamefont{Tolentino
  et~al.}(2004{\natexlab{b}})\citenamefont{Tolentino, Cezar, Souza-Neto,
  Tamura, Ramos, and Neueschwander}}]{Tolentino-DXAS04}
\bibinfo{author}{\bibfnamefont{H.~C.~N.} \bibnamefont{Tolentino}},
  \bibinfo{author}{\bibfnamefont{J.~C.} \bibnamefont{Cezar}},
  \bibinfo{author}{\bibfnamefont{N.~M.} \bibnamefont{Souza-Neto}},
  \bibinfo{author}{\bibfnamefont{E.}~\bibnamefont{Tamura}},
  \bibinfo{author}{\bibfnamefont{A.~Y.} \bibnamefont{Ramos}}, \bibnamefont{and}
  \bibinfo{author}{\bibfnamefont{R.}~\bibnamefont{Neueschwander}}
  (\bibinfo{year}{2004}{\natexlab{b}}), \bibinfo{note}{in preparation}.

\end{thebibliography}

\end{document}